\documentclass[fleqn,11pt]{article}

\usepackage{amsfonts,amsmath,amssymb}
\usepackage{latexsym}
\usepackage{orcidlink}
\usepackage{commath}
\hypersetup{colorlinks,citecolor=blue}
\hypersetup{colorlinks=true,linkcolor=red,filecolor=magenta, urlcolor=blue}
\usepackage{mathtools,amssymb,lipsum}
\usepackage{cuted}
\usepackage{etoolbox}
\usepackage{amsfonts,amssymb,cite}
\usepackage{graphicx}

\def\noi{\noindent}
\def\jnumber#1#2{\thispagestyle{empty} \noi\unitlength=1mm
    	\begin{picture}(178,10)
            \put(177,15){\llap{\large\it Grav. Cosmol. No.\,#1, #2}}
                    \end{picture}}

\newcommand{\Title}[1]{\noi {{\Large\bf #1}}\\[1ex]}

\def\Aunames#1{\noi{\bf #1}}
\def\au#1{${}^{#1}$}
\def\Addresses#1{\medskip\noi \protect
	\begin{description}\itemsep -3pt {\it #1} \end{description}}
\def\adr#1#2{\item[${}^{#1}$]{\it #2}}

\newcommand{\Abstract}[1]{\vskip 2mm \begin{center}
        \parbox{16.4cm}{\small\noi #1} \end{center}\medskip}

\def\email#1#2{\footnotetext[#1]{e-mail: #2}\addtocounter{footnote}{1}}

\def\nqq{\hspace*{-2em}}

\def\cm{\hspace*{1cm}}

\usepackage{color}

\def\Jl#1#2{#1 {\bf #2},\ }

\def\ApJ#1 {\Jl{Astroph. J.}{#1}}
\def\CQG#1 {\Jl{Class. Quantum Grav.}{#1}}
\def\DAN#1 {\Jl{Dokl. AN SSSR}{#1}}
\def\GC#1 {\Jl{Grav. Cosmol.}{#1}}
\def\GRG#1 {\Jl{Gen. Rel. Grav.}{#1}}
\def\IJMPD#1 {\Jl{Int. J. Mod. Phys. D}{#1}}
\def\JETF#1 {\Jl{Zh. Eksp. Teor. Fiz.}{#1}}
\def\JETP#1 {\Jl{Sov. Phys. JETP}{#1}}
\def\JHEP#1 {\Jl{JHEP}{#1}}
\def\JMP#1 {\Jl{J. Math. Phys.}{#1}}
\def\NPB#1 {\Jl{Nucl. Phys. B}{#1}}
\def\NP#1 {\Jl{Nucl. Phys.}{#1}}
\def\PLA#1 {\Jl{Phys. Lett. A}{#1}}
\def\PLB#1 {\Jl{Phys. Lett. B}{#1}}
\def\PRD#1 {\Jl{Phys. Rev. D}{#1}}
\def\PRL#1 {\Jl{Phys. Rev. Lett.}{#1}}

\def\lal{&&\nqq {}}

\def\beq{\begin{equation}}
\def\eeq{\end{equation}}
\def\bear{\begin{eqnarray}}
\def\bearr{\begin{eqnarray} \lal}
\def\ear{\end{eqnarray}}
\def\earn{\nonumber \end{eqnarray}}

\begin{document}
\twocolumn[
\jnumber{issue}{year}

\Title{Cosmological Model with Cosmic transit behaviour in Brans-Dicke Theory}
	   		
%\Author{First Author\au{1} and Second Author\au{2}}
 	     %{Affiliation }  	     
 	     %(The command ``\Author'' is used if the affiliation is the same for all authors)}
 	     
%\red{\bf or}  \bigskip     

\Aunames{Sunil K. Tripathy\au{a,1}, Alaka Priyadarsini Sendha\au{a,2}, Sasmita Kumari Pradhan\au{b,3}, Zashmir Naik\au{c,4}, B. Mishra\au{d,5}}
		%(The command ``\Aunames'' always together with ``\Addresses'', is used 
		%  if affiliations are different for different authors)

\Addresses{
\adr a {Department of Physics, Indira Gandhi Institute of Technology, Sarang, Dhenkanal, Odisha-759146, \\India.}
\adr b {Department of Physics, Centurion University of Technology and Management, Odisha, India and School of Physics, Sambalpur University, Jyotivihar, Sambalpur, Odisha-768019, India.}
\adr c {School of Physics, Sambalpur University, Jyotivihar, Sambalpur, Odisha, India, 768019.}
\adr d {Department of Mathematics, Birla Institute of Technology and Science-Pilani, Hyderabad Campus, Hyderabad-500078, India.}
}

%\bigskip

\Abstract
	{We have constructed a dark energy cosmological model in the framework of generalized Brans Dicke theory (GBD) with a self-interacting potential. The source of dark energy is considered through  a unified dark fluid (UDF) characterized by a linear equation of state (EoS). A time varying deceleration parameter simulating the cosmic transit behaviour has been introduced to get the dynamical behaviour of the model. The $H(z)$ data have been explored to constrain the model parameters and to study the dynamical aspects of the Brans-Dicke parameter and the scalar field.\\
	
	}
\medskip

] %%%%%%%%%%%%%%%%%%%%%%%%%%%%%%%%
\email 1 {tripathy\_sunil@rediffmail.com}
\email 2 {tripathy\_sunil@rediffmail.com}
\email 3 {sasmita.gita91@gmail.com}
\email 4 {zashmir@gmail.com}
\email 5 {bivu@hyderabad.bits-pilani.ac.in\\ \cm (Corresponding author)}
% ===============
\section{Introduction}
Cosmological observations have shown that the late time cosmic acceleration issue is due to the presence of an unknown form of energy, known as dark energy (DE) which occupies $68.3$ percent of the total mass energy budget of the Universe. Within the framework of GR, the explanation of DE requires some additional dynamical degrees of freedom in the form of scalar field. Even though, there have been geometrical modification routes suggested in literature to understand the late time cosmic speed up issue\cite{Carroll04,Nojiri05,Nojiri07,Linder10,Harko11}, the scalar field mediated gravity theories have gained much importance. The scalar fields provide dynamical variables in the gravitational field equations and also be useful in studying the early evolutionary or inflationary era.  Among the scalar-tensor theories proposed, Brans-Dicke (BD) theory \cite{Brans61} is considered to be most successful. In the BD theory, the scalar field ($\phi$) is coupled to the geometry with a coupling constant known as BD parameter ($\omega$). In this process of coupling, the gravity is mediated through the scalar field and we naturally get a time varying Newtonian Gravitational constant ($G$)as an inverse of the scalar field. The BD theory has passed the experimental test from solar system \cite{Bertotti03} and is also tested against the Cosmic Microwave Background (CMB) data and large scale structure \cite{Xu10, Xu13}. Scalar-tensor theories of gravity including the BD theory have been used to model the variations of constants of nature including $G$ \cite{Uzan2003, Uzan2011}. Also they provide the archetypal class of theories used to quantify deviations from Einstein’s theory\cite{Will93, Durk2019}.

The study of DE cosmological models is the emerging research in modern cosmology. Eingorn and Kiefer \cite{Eingorn15} have investigated the scalar cosmological perturbations in the framework of a model with interacting DE and dark matter. Xu and Zhang \cite{Xu16} have compared ten popular DE models according to their capabilities of fitting the current observational data. Zhai et al. \cite{Zhai17} compared several theoretical cosmological models with the  observational data from the Type Ia supernova measurements of expansion, cosmic microwave background, baryon acoustic oscillation measurements of expansion and the local Hubble constant. Bruck and Mifsud \cite{Bruck18} have considered cosmological models that allow  non-gravitational direct couplings between dark matter and DE and distinguished the cosmological features of these couplings by the cosmological observations. Mishra et al. \cite{Mishra19} and Ray et al. \cite{Ray19} studied the two fluid cosmological models in anisotropic space-time with the energy momentum tensor as the combination of DE fluid and ordinary matter fluid.  Without any field theoretic interpretation, Paul and Sengupta \cite{Paul20} have explained the recent observations with a dynamical cosmological constant. Akarsu et al. \cite{Akarsu20} have studied the cosmological constant in the standard $\Lambda$CDM model by introducing the graduated DE. Several research work have been performed in recent times on the issue of late time cosmic speed up phenomena and the associated exotic dark energy concept with the scalar-tensor BD theory that provide some ample insight into the accelerated expansion. Banerjee and Pavon \cite{Banerjee01} have argued that BD theory can explain the late time cosmic expansion of the Universe without resorting the cosmological constant or quintessence matter. Kim \cite{Kim05} suggested that BD theory as unified model for dark matter and DE. Acquaviva and Verde \cite{Acquaviva07} have framed the Jordan BD model of gravity and with some analysis shown the inconsistencies in the expansion history studied through the usual approach. In the framework of BD theory, Chattopadhyay et al. \cite{{Chattopadhyay14}} have used some reconstruction scheme
for new holographic DE model and derived the modified field equation of BD theory to study the expansion history of the universe. Sharif and Shah \cite{Sharif19} have discussed the stability of the pilgrim DE model in BD theory. Zadeh and Sheykhi \cite{Zadeh20} have studied the agegraphic dark energy in the background of BD theory with pressureless dark matter. Tripathy et al. \cite{Tripathy2020} have studied the cosmology with a hybrid scale factor in BD theory. Verma has analysed the data of gravitational waves from spinning neutron stars within Brans-Dicke theory \cite{Verma2021}. Du has studied the scalar stochastic gravitational wave background from compact binary mergers and stellar collapses in BD theory \cite{Du2019}.

The generalization of BD theory was proposed long time back to incorporate the various requirements for the value of the BD parameter, where $\omega$ is considered as a function of the scalar field $\phi$ \cite{Bergmann68,Nordtvedt70,Wagoner70}. The reason behind the generalization is that for $2\omega+3=\frac{1}{\alpha \phi}$ and $2\omega+3=\frac{1}{\phi-1}$, GBD reduces to  Schwinger’s theory\cite{Schwinger70} and to Barker's scalar tensor theory \cite{Barker78} respectively. However, the controversy in estimating the value of the BD parameter $\omega$ continues. Several experiments have suggested the range for the BD parameter, prominent among them are solar system experiment, $\omega>40000$ \cite{Bertotti03}; WMAP and SDSS, $\omega<-120$ \cite{Wu10,Wu13}; $417<\omega<714$ \cite{Alavirad14}. In a recent work, within the framework of GBD theory, Tripathy et al. \cite{Tripathy15} have shown the linkage between the rate of anisotropy and the non-evolving part of the BD parameter. In another work, Tripathy et al.\cite{Tripathy2020} have obtained low negative values for the BD parameter in the range $\omega>-1.5$. Low negative value of the BD parameter is required to explain the structure formation, coincidence problem and late time cosmic acceleration \cite{Sahoo2002, Bert2000, Banerjee2001, Sen2001a, Sen2003}. Pradhan et al. have discussed the possibility of a Big Rip scenario within the generalized BD theory and have shown that the Universe within the specified model may witness a finite time doomsday at a time $t_{BR}\approx 16.14 GYr$ requiring a large negative value of the BD parameter \cite{Pradhan2022}.  Because of the success of the GBD theory in explaining the key issues in cosmology, recently there have an increasing interest in this scalar-tensor theory either in the original form or with geometrical modification of the Einstein-Hilbert action \cite{Roy17, Zucca20,Sola2020,Bonino2020, Leon2020}.

In the present work, we wish to construct some dark energy cosmological models within the framework of the generalized Brans-Dicke theory. Since, the present observations regarding the late time cosmic speed up issue hint for a cosmic transit from a decelerated phase of expansion to an accelerated phase, we will construct such a cosmic transit model which may be simulated by a signature flipping deceleration parameter. In fact, the redshift $z_{t}$ at which the transition occurs may be considered as an important parameter and is order of unity i.e $z_{t} \sim 1$.  There have been different constraints on the cosmic transit redshift from observational data. Some of recent constraints on the transition redshift parameter may include Farooq and Ratra:  $z_{t}=0.74\pm 0.05$ \cite{Farooq13}, Capozziello et al.: $z_{t}= 0.7679^{+0.1831}_{-0.1829}$ \cite{Capo14},  Reiss et al.: $z_{t}= 0.426^{+0.27}_{-0.089}$ \cite{Reiss07}, Lu et al. : $z_{t}=0.69^{+0.23}_{-0.12}$ \cite{Lu11}, Moresco et al.:$z_{t}=0.4\pm 0.1$ \cite{Moresco16}, Yang and Gong: $z_{t}=0.61^{+0.231}_{-0.12}$ \cite{Yang2019},  Jesus et al.: $z_{t}=0.806\pm 0.094$ \cite{Jesus2018}. Cosmological models showing cosmic transit behaviour with proper prediction of the transition redshift may be considered to be more viable and realistic. We use the observational $H(z)$ data to the constrain the model parameters and to study the cosmic transit behaviour. The paper is organized as: In Sec. II, the basic field equations with  a time varying BD parameter and self-interacting potential are derived. The DE EoS is presented in Sec. III. In Sec. IV, the cosmic transit behaviour along with the variable deceleration parameter are analysed. The behaviour of the time varying BD parameter and the BD scalar field are discussed. The final remarks along with the conclusions of the present work are listed in Sec. V.

\section{Basic Formalism}\label{II}
The action for generalized Brans-Dicke (GBD) theory in a Jordan frame is given by \cite{Bergmann68,Nordtvedt70,Wagoner70} 
\begin{equation}
S=\int d^{4}x\sqrt{-g}\left[\varphi R - \frac{\omega(\varphi)}{\varphi}\varphi^{,\alpha}\varphi_{,\alpha}-V(\varphi)+\mathcal{L}_m\right],\label{eq:1}
\end{equation}
where $\omega(\varphi)$ is the modified Brans-Dicke parameter that varies with the Brans-Dicke scalar field $\varphi$. $V(\varphi)$ is the self-interacting potential, $R$ is the scalar curvature and $\mathcal{L}_m$ is the matter Lagrangian. The field equations of GBD theory can be obtained as  
\begin{eqnarray}
G_{\mu\nu} - \dfrac{\omega(\varphi)}{\varphi^{2}}\left[\varphi_{,\mu}\varphi_{,\nu}-\frac{1}{2}g_{\mu\nu}\varphi_{,\alpha}\varphi^{,\alpha}\right]\nonumber\\-\frac{1}{\varphi}[\varphi_{,\mu;\nu}-g_{\mu\nu}\Box\varphi]+\frac{V(\varphi)}{2\varphi}g_{\mu\nu}&=& \frac{T_{\mu\nu}}{\varphi},\label{eq:2}
\end{eqnarray}
\begin{eqnarray}
\Box\varphi &=&\frac{1}{2\omega(\varphi)+3} \Box[T-\left(2V(\varphi)-\varphi\frac{\partial V(\varphi)}{\partial\varphi}\right)\nonumber\\& & -\frac{\partial\omega(\varphi)}{\partial\varphi}\varphi_{,\mu}\varphi^{,\mu}\Box].\label{eq:3}
\end{eqnarray}
Here we chose $8\pi G_0=c=1$; $G_0$ and $c$ respectively denote the Newtonian Gravitational Constant and the speed of light in vacuum. $T=g^{\mu\nu}T_{\mu\nu}$ is the trace of the energy momentum tensor $T_{\mu\nu}$, $\Box$ is the d'Alembert operator. In the present work, we assume the energy momentum tensor as $T_{\mu\nu}=(\rho+p)u_{\mu}u_{\nu}+pg_{\mu\nu}$ where $\rho$ is the dark energy density  and $p$ is the dark energy pressure. We consider a plane symmetric LRS Bianchi I (LRSBI) metric in the form
\begin{equation}
ds^2=-dt^2+a_1^2dx^2+a_2^2(dy^2+dz^2), \label{eq:4}
\end{equation}
where $a_1$ and $a_2$ are functions of cosmic time only. The GBD field equations for the anisotropic LRSBI Universe can be obtained as, 
\begin{center}
\begin{strip}
\begin{eqnarray}
2\frac{\dot{a}_1\dot{a}_2}{a_1a_2} +\left(\frac
{\dot{a}_2}{a_2}\right)^2 -\frac{\omega(\varphi)}{2}\left(\frac{\dot{\varphi}}{\varphi}\right)^{2}+\left(\frac{\dot{a}_1}{a_1}+2\frac{\dot{a}_2}{a_2}\right)\frac{\dot{\varphi}}{\varphi}-\frac{V(\varphi)}{2\varphi} &=& \frac{\rho}{\varphi},\label{eq:5}\\
2\frac{\ddot{a}_2}{a_2}+\left(\frac{\dot{a}_2}{a_2}\right)^2+\frac{\omega(\varphi)}{2}\left(\frac{\dot{\varphi}}{\varphi}\right)^{2}+2\left(\frac{\dot{a}_2}{a_2}\right)\frac{\dot{\varphi}}{\varphi}+\dfrac{\ddot{\varphi}}{\varphi}-\dfrac{V(\varphi)}{2\varphi} &=&- \frac{p}{\varphi}
,\label{eq:6}\\
\frac{\ddot{a}_1}{a_1}+\frac{\ddot{a}_2}{a_2}+\frac{\dot{a}_1\dot{a}_2}{a_1a_2}+\frac{\omega(\varphi)}{2}\left(\frac{\dot{\varphi}}{\varphi}\right)^{2}+\left(\frac{\dot{a}_1}{a_1}+\frac{\dot{a}_2}{a_2}\right)\frac{\dot{\varphi}}{\varphi}+\dfrac{\ddot{\varphi}}{\varphi}-\dfrac{V(\varphi)}{2\varphi} &=& -\frac{p}{\varphi}
.\label{eq:7}
\end{eqnarray}
\end{strip}
\end{center}
The directional Hubble parameters can be expressed as  $H_x=\frac{\dot{a_1}}{a_2}$ and $H_y=\frac{\dot{a_2}}{a_2}$. Incorporating an anisotropic relation among the directional Hubble parameters  as, $H_{x}=kH_{y}, k\neq 1$, the mean Hubble parameter may be expressed as $H=\frac{1}{\xi}H_{y}$, where the anisotropic parameter $\xi$ is related to the parameter $k$ as $\xi=\frac{3}{k+2}$. It is obvious that, for $\xi=1$, the model reduces to that of a flat isotropic model. In terms of the Hubble parameter and the anisotropic parameter $\xi$, the GBD field equations are expressed as \cite{Tripathy2020}

\begin{eqnarray}
(2k+1)\xi^2H^{2} &=&\frac{1}{\varphi}\left(\rho+\rho_{\varphi}\right), \label{eq:8}
\end{eqnarray}
\begin{eqnarray}
2\xi\dot{H}+3\xi^2H^{2} &=& -\frac{1}{\varphi}\left(p+p^1_{\varphi}\right),\label{eq:9}
\end{eqnarray}
\begin{eqnarray}
(k+1)\xi\dot{H}+(k^{2}+k+1) \xi^2 H^{2}&=& -\frac{1}{\varphi}\left(p+p^2_{\varphi}\right),\label{eq:10}\nonumber\\
\end{eqnarray}
where we have
\begin{eqnarray}
\rho_{\varphi} &=&\frac{1}{2}\omega(\varphi)\varphi\left(\frac{\dot{\varphi}}{\varphi}\right)^{2}-3H\dot{\varphi}+\frac{1}{2}V(\varphi), \label{eq:11}\\
p^1_{\varphi} &=&\frac{1}{2}\omega(\varphi)\varphi\left(\frac{\dot{\varphi}}{\varphi}\right)^{2}+2\xi H\dot{\varphi}+\ddot{\varphi}-\dfrac{1}{2}V(\varphi),\label{eq:12}\\
p^2_{\varphi}&=& \frac{1}{2}\omega(\varphi)\varphi\left(\frac{\dot{\varphi}}{\varphi}\right)^{2} +(k+1)\xi H\dot{\varphi}+\ddot{\varphi}-\frac{1}{2}V(\varphi).\label{eq:13}\nonumber\\
\end{eqnarray}

The pressure anisotropy for the DE pressure may now be expressed as
\begin{equation}
\triangle p_{\varphi}=p^2_{\varphi}-p^1_{\varphi}=(k-1)\xi H\dot{\varphi}.\label{eq:14}
\end{equation}
It is obvious that, the pressure anisotropy vanishes for an isotropic model. Also, we do not have a pressure anisotropy for a constant BD scalar field. The wave equation for the scalar field becomes
\begin{eqnarray}
\dfrac{\ddot{\varphi}}{\varphi}+3H\frac{\dot{\varphi}}{\varphi}= \frac{1}{{2\omega(\varphi)+3}}\Box[\rho-3{p}-\frac{\partial\omega(\varphi)}{\partial\varphi}\dot{\varphi}^{2}\nonumber\\ -2V(\varphi)+\varphi\frac{\partial V(\varphi)}{\partial\varphi}\Box]. \label{eq:15}
\end{eqnarray}

The scalar field dependence of the Brans-Dicke parameter $\omega(\varphi)$ and self-interacting potential $V(\varphi)$ may be obtained from the Eqs. \eqref{eq:8}-\eqref{eq:10} as \cite{Tripathy2020, Pradhan2022},
\begin{eqnarray}
\omega(\varphi) &=& \left(\frac{\dot{\varphi}}{\varphi}\right)^{-2}\Box[-\dfrac{\rho+p}{\varphi}-\frac{\ddot{\varphi}}{\varphi}+k\xi H\dfrac{\dot{\varphi}}{\varphi}\nonumber\\ & &-2\xi \dot{H}-2(1-k)\xi^2 H^{2}\Box],\label{eq:16}
\end{eqnarray}
\begin{eqnarray}
V(\varphi) &=& 2\xi^2H^2\varphi-2\rho-\omega(\varphi)\varphi\left(\frac{\dot{\varphi}}{\varphi}\right)^{2}+6H\dot{\varphi}. \label{eq:17}\nonumber\\
\end{eqnarray}

The dynamical behaviour of the BD parameter and the self interacting potential requires the knowledge of the equation of state in the form of $p=p(\rho)$ and the dynamical evolution of the scale factor. In the present work, we are interested in a cosmic transit model that simulates a signature flipping behaviour of the deceleration parameter which in turn fixes the Hubble rate. From a tabulated observational $H(z)$ data we will fix the scale factor. Another aspect is the fixation of the anisotropic parameter $\xi$. In fact, recent works on cosmic anisotropy show a small value for the cosmic shear which provides a little deviation of the anisotropic parameter from its isotropic value $1$. The cosmic anisotropy issue has not yet resolved substantially and therefore, we intend to keep this factor considering a little departure from unity. 
\section{Equation of state}

In order to handle the dark energy issue, we consider a linear relationship between the pressure and energy density in the form 
\begin{equation} \label{eq:18}
p=\zeta(\rho-\rho_{0}),
\end{equation} 
where $\zeta$ and $\rho_{0}$ are positive constants. A hydrodynamically stable fluid description may be obtained for a positive value of $\zeta$ \cite{Babichev2004}. The choice of the linear equation of state speculates an approach to unify the dark matter and dark energy into a single dark fluid known as unified dark fluid (UDF) \cite{Tripathy15}. In fact, the UDF EoS \eqref{eq:18} signifies the degeneracy of the dark sector as its first part behaving like a usual cosmic fluid and the other as a cosmological constant.  Moreover, $\zeta=0$ refers to the case of dark matter and $\zeta=1$ implies a stiff fluid dominated with dark energy. In literature, there have been many suggestions on the generalized EoSs where the dark energy and dark matter are considered to be two different aspects of the same fluid \cite{Capozziello06,Liao12, Xu12}. In some of these unified dark sector models, the EoS may depend on the energy density, the Hubble parameter and its derivatives \cite{Capozziello06}.

The conservation equation for the given UDF may be expressed as
\begin{equation}\label{eq:19}
\dot{\rho}+3H\left[\kappa\left(\rho-\rho_0\right)+\rho_0\right]=0,
\end{equation}
which may be integrated to obtain the energy density as
\begin{equation} \label{eq:20}
{\rho}={\rho_{\Lambda}}+\rho_{\kappa}\left(\frac{\mathcal{R}}{\mathcal{R}_0}\right)^{-3\kappa},
\end{equation}
where, $\rho_{\Lambda}= \left(1-\frac{1}{\kappa}\right) \rho_{0}$,  $\rho_{\kappa}=\rho_1-\rho_{\Lambda}$ and $\kappa=\zeta+1$. $\rho_{1}$ is related to the dark energy density at the present epoch and $\mathcal{R}$ is the scale factor.  $\mathcal{R}_0$ is the scale factor at the present epoch. The dark energy pressure becomes
\begin{equation}\label{eq:21}
p={-\rho_{\Lambda}}+\left(\kappa-1\right)\rho_{\kappa}\left(\frac{\mathcal{R}}{\mathcal{R}_0}\right)^{-3\kappa}.
\end{equation}
The energy density and pressure for the given UDF equation of state may be expressed in terms of the redshift $z=-1+\left(\frac{\mathcal{R}}{\mathcal{R}_0}\right)^{-1}$ as 
\begin{eqnarray}
\rho &=& {\rho_{\Lambda}}+{\rho_{\kappa}}(1+z)^{3\kappa},\\ \label{eq:22}
p &=& {-\rho_{\Lambda}}+\left(\kappa-1\right)\rho_{\kappa}(1+z)^{3\kappa}. \label{eq:23}
\end{eqnarray}

At an early epoch, we have $z\rightarrow \infty$ and consequently, the magnitudes of the energy density and pressure become infinitely large. On the other hand, with the cosmic evolution, as we move to late phase, $z\rightarrow -1$, the energy density overlaps with $\rho_{\Lambda}$ and the pressure becomes $p=-\rho_{\Lambda}$. In view of this, we may identify $\rho_{\Lambda}$ as the vacuum energy density. At the present epoch, ($z=0$), these quantities respectively assume the values $\rho=\rho_{\Lambda}+\rho_{\kappa}$ and $p=-\rho_{\Lambda}+(\kappa-1)\rho_{\kappa}$. As has been supported by observational facts, in the present model, the late phase of the cosmic evolution witnesses an accelerated expansion mostly dominated by the dark energy. 

Since, for the present UDF model, we have $\rho+p=\kappa \rho_{\kappa}(1+z)^{3\kappa}$, the dark energy EoS, $\omega_{D}=-1+\frac{\rho+p}{\rho}$ may be obtained as

\begin{equation}\label{eq:24}
\omega_{D}=-1+\dfrac{\kappa}{1+\left(\dfrac{\rho_{\Lambda}}{\rho_{\kappa}}\right)(1+z)^{-3\kappa}},
\end{equation}
which may also be expressed as
\begin{equation}\label{eq:25}
\omega_{D}=-1+\dfrac{\kappa}{1+\frac{\chi}{1-\chi}(1+z)^{-3\kappa}},
\end{equation}
where $\chi=\left(\dfrac{\rho_{\Lambda}}{\rho_{1}}\right)$. In fact the dark energy density parameter $\Omega_D=\frac{\rho_{\Lambda}}{3H^2}\simeq 0.7$ and it dominates the energy density from usual matter at the present epoch. In view of this, we may infer that, $\frac{\rho_{\Lambda}}{\rho_{\kappa}}>>1$. In some of the earlier work, attempts have been made to constrain the UDF model parameters from reasonable basis \cite{Liao12, Tripathy15, Babichev2004, Tripathy2020, Tripathy20a}. One may note from the above expression (21) that, at an early epoch ($z\rightarrow \infty$), $\omega_D \rightarrow -1+\kappa$ and at a late epoch ($z\rightarrow -1$), $\omega_D \rightarrow -1$. In otherwords, the present UDF model overlaps with the concordant $\Lambda$CDM model at late times and  is in conformity with recent observations. At the present epoch, we may have approximate value $\omega_D \simeq -1 -\kappa\chi$. This approximate relation may be used to constrain the values of the parameter $\kappa$ from the known observational constraints on the dark energy EoS. For an assumed value of  $\chi$, we may obtain $\kappa$ from known $\omega_D$ through the relation
\begin{equation}\label{eq:26}
\kappa=-\frac{1+\omega_D}{\chi}.
\end{equation}

Since $\chi$ is assumed to be a positive quantity, a positive $\kappa$ requires that $\omega_D<-1$ which shows that, the present UDF model may favour a phantom field dominated universe. In an earlier work \cite{Tripathy2020}, similar conclusion has been derived for the UDF model within a GBD theory using  experimental $H(z)$ data.

\section{Variable Deceleration Parameter And Cosmic Transit Behaviour}
It is confirmed from a lot of observations in recent times that, the universe is speeding up in the present epoch \cite{Riess98,Perlmutter99}. It is believed that the Universe may have undergone a transition from a decelerated phase to an accelerated one. The cosmic redshift at which this transition occurs is called transition redshift $z_{t}$. This implies a signature flipping  of the deceleration parameter $q$ from a positive value at some early time to negative value at late time of cosmic evolution. Here, we will construct some cosmological models that are consistent with this behaviour of the Universe and can provide a signature flipping of the deceleration parameter. We consider a scale factor that simulates a cosmic transit behaviour from an early decelerating phase to a late time acceleration as,

\begin{equation} \label{eq:27}
\mathcal{R}=\left(\frac{e^{\delta\gamma{t}}-\beta}{\delta}\right)^\frac{1}{\gamma},
\end{equation}
where $\delta, \beta$ and $\gamma$ are the model parameters. The Hubble parameter for this model becomes $ H=\delta+\frac{\beta}{{\mathcal{R}}^{\gamma}}$ and can be expressed in terms of redshift $z$ as

\begin{equation}
H(z)=\delta+\beta\left(1+z\right)^\gamma. \label{eq:28}
\end{equation}
Here we have defined the redshift as $1+z=\frac{1}{\mathcal{R}}$ by assuming the scale factor at the present epoch to be unity i.e $\mathcal{R}_0=1$. the scale factor at the present epoch to be $1$. For the present model, the deceleration parameter is obtained as,
\begin{equation}\label{eq:29}
q=-1-\frac{\dot{H}}{H^2} =-1+\frac{\beta\gamma}{\delta{\mathcal{R}}^\gamma+\beta}.
\end{equation}
One may infer that, at an initial epoch with $\mathcal{R}\rightarrow 0$, the deceleration parameter is $q=\gamma-1$. At a late phase of cosmic evolution with $\mathcal{R}\rightarrow \infty$, we have $q=-1$. At the present epoch, it becomes $q_0=-1+\frac{\beta\gamma}{\delta+\beta}$. It is clear that, for $\gamma>1$, the deceleration parameter evolves from a positive value at an early epoch to overlap with the concordance $\Lambda$CDM value and/or with the de Sitter kind of expansion at late phase.

\subsection{Transition redshift} 
On the basis of a lot of recent observational data, the Universe is believed to have undergone a transition from deceleration to acceleration phases at a transition redshift $z_{t}$ whose value primarily depends on the cosmology and the assumed gravitational theory. $z_{t}$ is of the order of unity i.e. $z_{t} \sim 1$. In the present work, we considered the available observational $H(z)$ data to constrain the model parameters through a $\chi^{2}$ fitting procedure 
\begin{equation} \label{eq:30}
{\chi^{2}_H}(p)={\sum^N_{i=1}}\frac{(H^t-H^o)^{2}}{\sigma^{2}_H}
\end{equation}
for $N$ measured values of $ H^o$ with variance $\sigma^{2}_H$ at red shift ${z}$ where $H^{t}$ is the predicated value of $H(z)$ from the present cosmological model. In the fitting procedure, the constrained value of the parameters are obtained as $\delta=36.67\pm 9.58$, $\beta=27.90\pm 7.09$, $\gamma=1.58\pm 0.17$. The $\chi^{2}$ is found to be $19.62$. In FIG. 1, the  Hubble parameter $H(z)$ for the present model with the constrained parameter space is shown as a function of the redshift.  The observational data for the Hubble parameter are also shown in the figure as unconnected solid dots. In FIG.2, $\frac{H(z)}{1+z}=\dot{\mathcal{R}}$ is plotted as a function of the redshift along with the observational values. This function $H(z)$ directly maps the accelerating behaviour of the Universe. For accelerating Universe, it has a positive slope and for a decelerating Universe, it should have a negative slope. Since the plot shows a positive slope for $H(z)$, the accelerating nature of the present Universe is quite certain which is also supported by the behaviour of the acceleration plot for $\frac{H(z)}{1+z}=\dot{\mathcal{R}}$ in FIG.2.

\begin{figure}[!htp]
\centering
\includegraphics[scale=0.35]{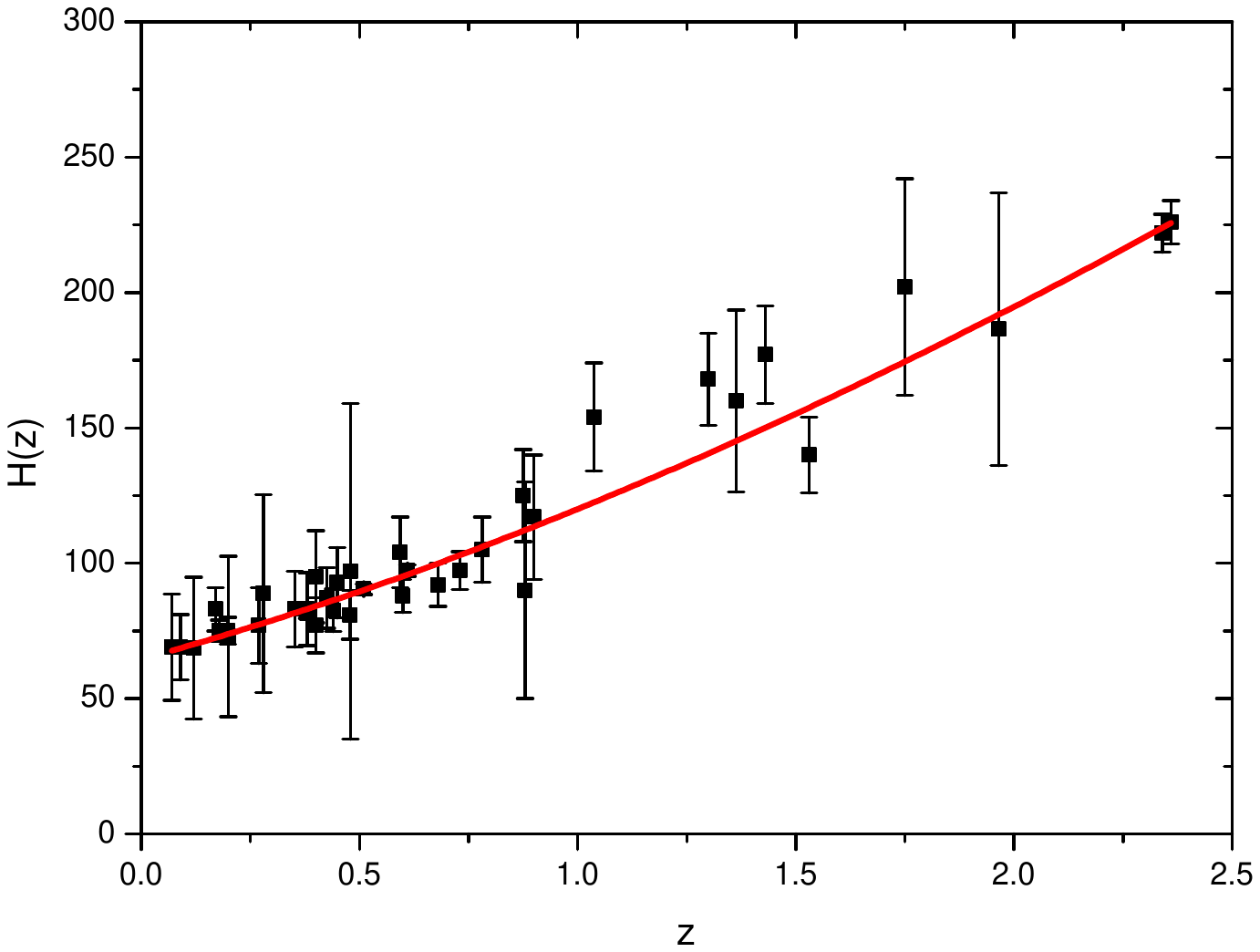}
\caption{ Hubble parameter $H(z)$ as function of redshift $z$ with unit $Kms^{-1}Mpc^{-1}$.}
\label{Fig1}
\end{figure}

\begin{figure}[!htp]
\centering
\includegraphics[scale=0.35]{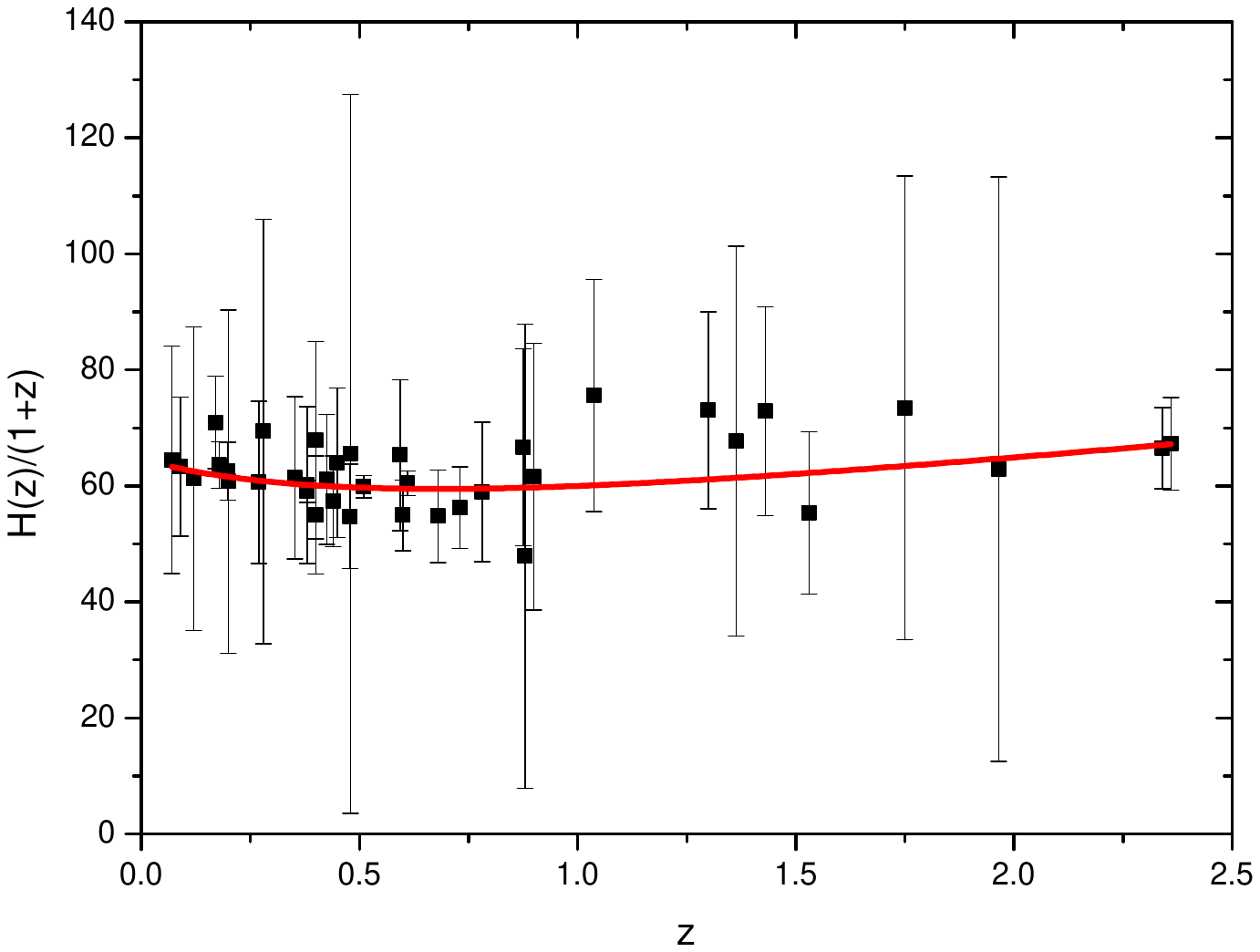}
\caption{  $\frac{H(z)}{1+z}$ as a function of redshift $z$ with unit $Kms^{-1}Mpc^{-1}$ for the constructed models.}
\label{Fig2}
\end{figure}

%*********************************
%*********************************
\begin{figure}[!htp]
\centering
\includegraphics[scale=0.35]{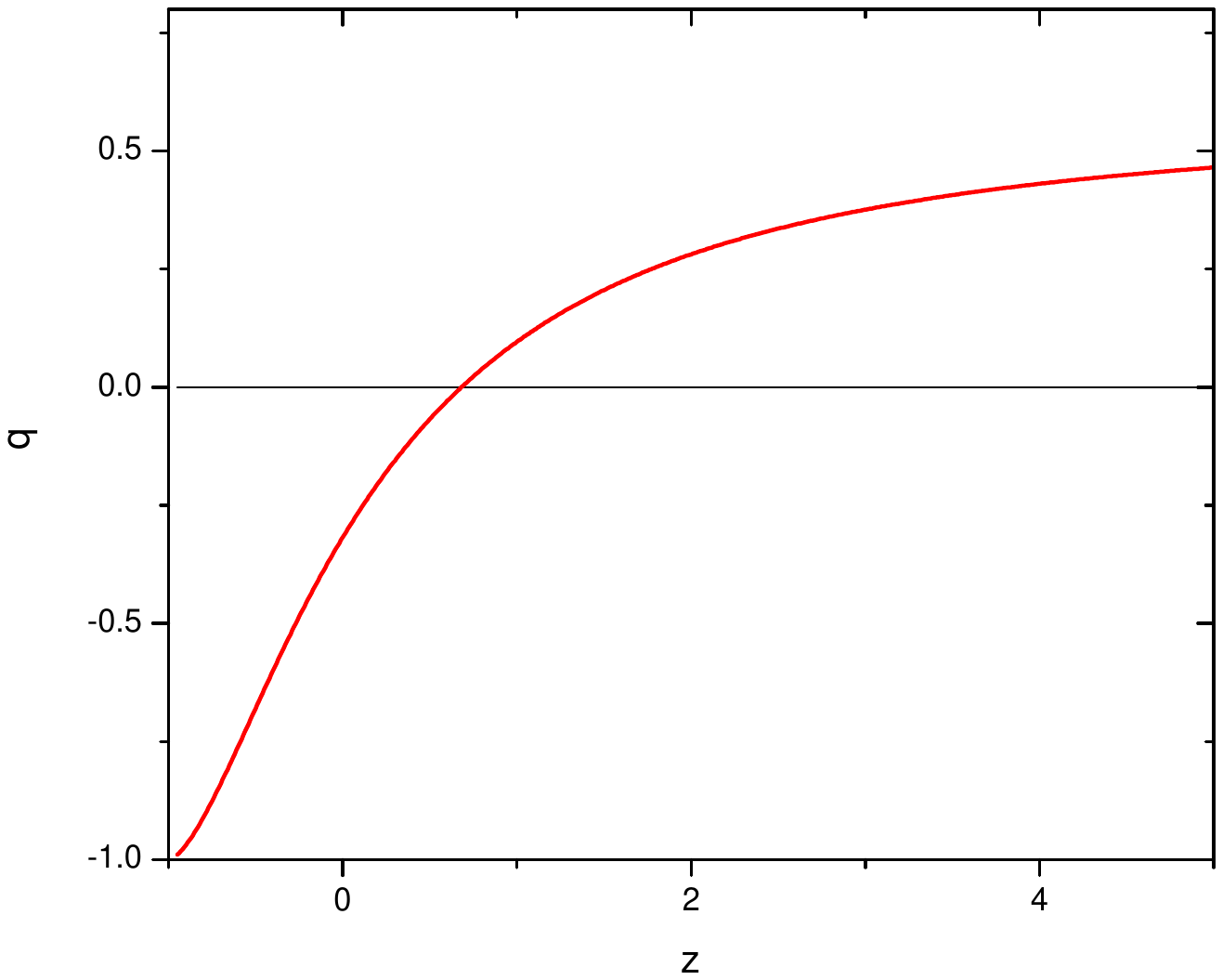}
\caption{ Deceleration parameter $q$ as a function of redshift $z$ for the constructed models.}
\label{Fig3}
\end{figure}
%*********************************
%*********************************
The so called $H_0$ tension arises out of the discrepancy between the local distance ladder measurement  and those inferred from Cosmic Microwave Background Radiation (CMBR) poses question on the the success of $\Lambda$CDM model\cite{Abdalla2022,Riess2016, Riess2018, Aghanim2020}. While the measurement of Riess et al. provides $H_0=74.03\pm 1.42 km~s^{-1}Mpc^{-1}$ \cite{Riess2016, Riess2018, Riess2022}, the CMB temperature from Planck collaboration suggests $H_0=67.36\pm 0.54 km~s^{-1}Mpc^{-1}$ \cite{Aghanim2020}. On the other hand, the results from the cosmological ladder and classical Cepheids and  supernovae type Ia of the S$H0$ES collaboration estimates $H_0=73.3^{+1.7}_{-1.8} km~s^{-1}Mpc^{-1}$ \cite{Reid2019} and the H0LiCOW collaboration tells us that $H_0=73.3^{+1.7}_{-1.8} km~s^{-1}Mpc^{-1}$\cite{Wong2019}. Eventhough the methods of determination of the value of the Hubble parameter at the present epoch have substantial accuracy, within the standard $\Lambda$CDM model, both the approaches do not reconcile and there emerges a tension of $5.0 \sigma$.  Till date there is no satisfactory explanation for the tension arising on the systematic errors. There are many attempts in recent times to reduce the $H_0$ tension. Recently, it has been shown that, the BD theory coupled with a $\Lambda$-term or a modified gravitational action may be helpful in reducing the above mentioned tensions \cite{Lambiase2019, Sola2020}. In a recent work, Capozziello et al. have considered the look back time  and inferred that the Hubble tension can be removed if the look-back time is correctly referred to the redshift where the measurement is performed \cite{Capo2023}. The discrepancy in the measurements from different observations leading to the $H_0$ tension may hint for a new Physics involving the dark energy and/or dark matter components either beyond the standard $\Lambda$CDM model, or beyond the standard model of particle physics \cite{Bernal2016,Mortsell2018}. As reported in \cite{Capo2023}, different estimates of the Hubble parameter from different measurements include $H_0=69.8\pm 0.8 km~s^{-1}Mpc^{-1}$ \cite{Freedman2019}, $H_0=67.8\pm 1.3 km~s^{-1}Mpc^{-1}$ \cite{Macaulay2019} and  $H_0=69.5^ {+3.0}_{-3.5} km~s^{-1}Mpc^{-1}$ \cite{Farren2022}. Goswami et al. have constrained the Hubble parameter from the analysis of joint $H(z)$ and Pantheon data as $H_0=69.36\pm 1.42 km~s^{-1}Mpc^{-1}$ \cite{Goswami2021}. In the present work, through the procedure of constraining the values of the model parameters, we obtain the Hubble parameter at the present epoch as $H_0=64.57\pm 16.67 km~s^{-1}Mpc^{-1}$ which is well within the recent estimates.\\

The signature flipping of the deceleration parameter $q$ for the constructed model is shown in In FIG-3. As we have discussed earlier, the deceleration parameter smoothly evolves from positive values in initial epoch to become $q=-1$ at late times showing a transition from a decelerated phase of expansion to an accelerated phase. The transition redshift corresponding to $q=0$ is obtained as $z_t=0.68$. This value of transition redshift is in conformity with that constrained from different analysis \cite{Capo14,Reiss07,Lu11,Yang2019,Goswami2021}. The deceleration parameter at the present epoch is obtained in the present analysis as $q_0 (z=0)=-0.32\pm 0.42$ signifying an accelerating phase of expansion at the present time.

\subsection{Brans-Dicke scalar field}
From the field equations \eqref{eq:6} and \eqref{eq:7}, one may get a relation between the BD scalar field and the Hubble parameter  as $\frac{\dot{\varphi}}{\varphi}=-\frac{\dot{H}}{H}-3H$ and on integration, we obtain 
\begin{equation} \label{eq:31}
\varphi= \varphi_{0}\left[\frac{H_{0}}{H{\mathcal{R}^3}}\right],
\end{equation}
where, $\varphi_{0}$ is the value of the scalar field at the present epoch. The scalar field depend on the scale factor and decreases with the growth of scale factor. The evolution of the BD scalar field is presented graphically in FIG. 4 as a function of redshift. The scalar field, for the constructed model, decreases from some large values at early phase to small values at the present epoch.
\begin{figure}[!t]
\centering
\includegraphics[scale=0.35]{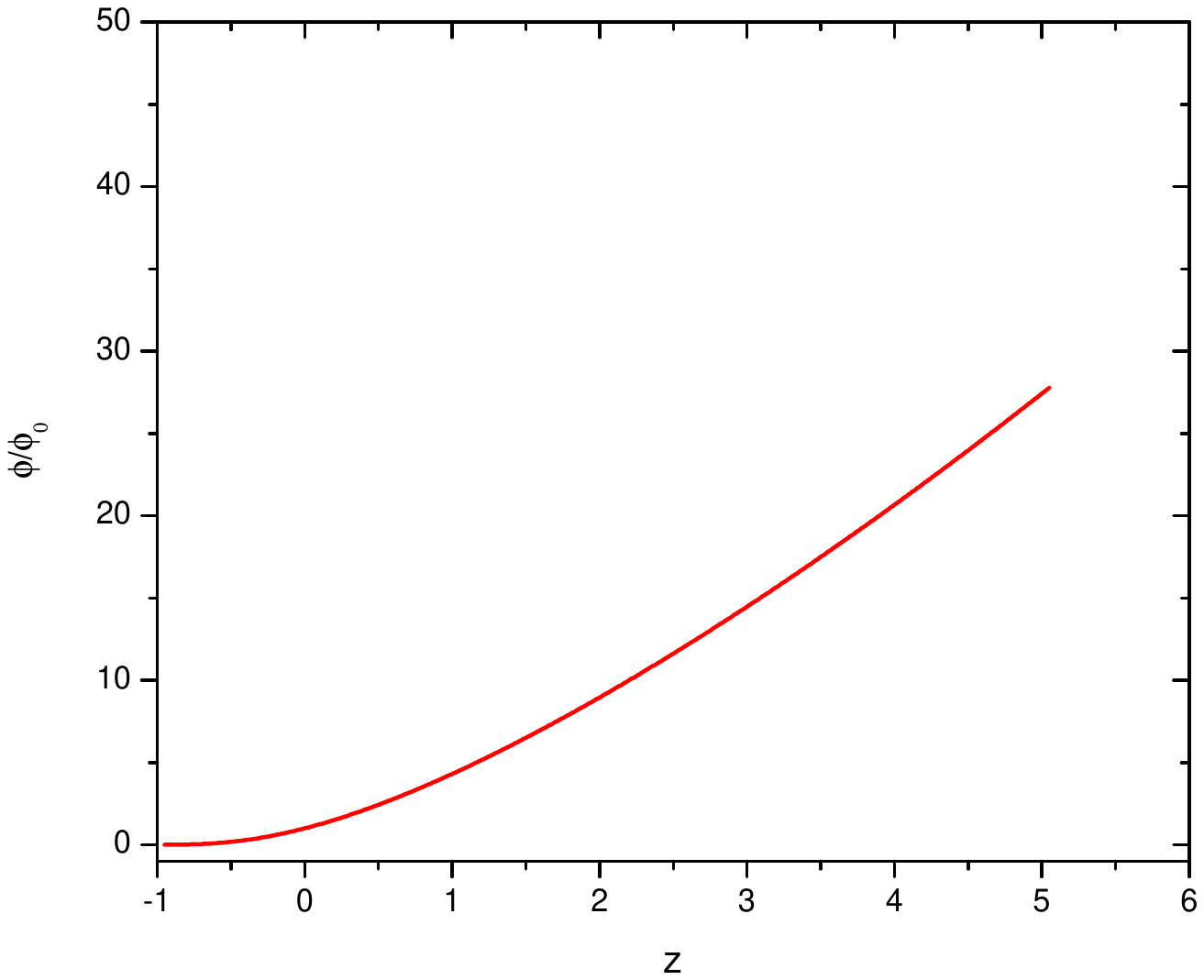}
\caption{ Evolution of the Brans-Dicke scalar field $\varphi$ with red shift $z$.}
\label{Fig4}
\end{figure}
\begin{figure}[!htp]
\centering
\includegraphics[scale=0.35]{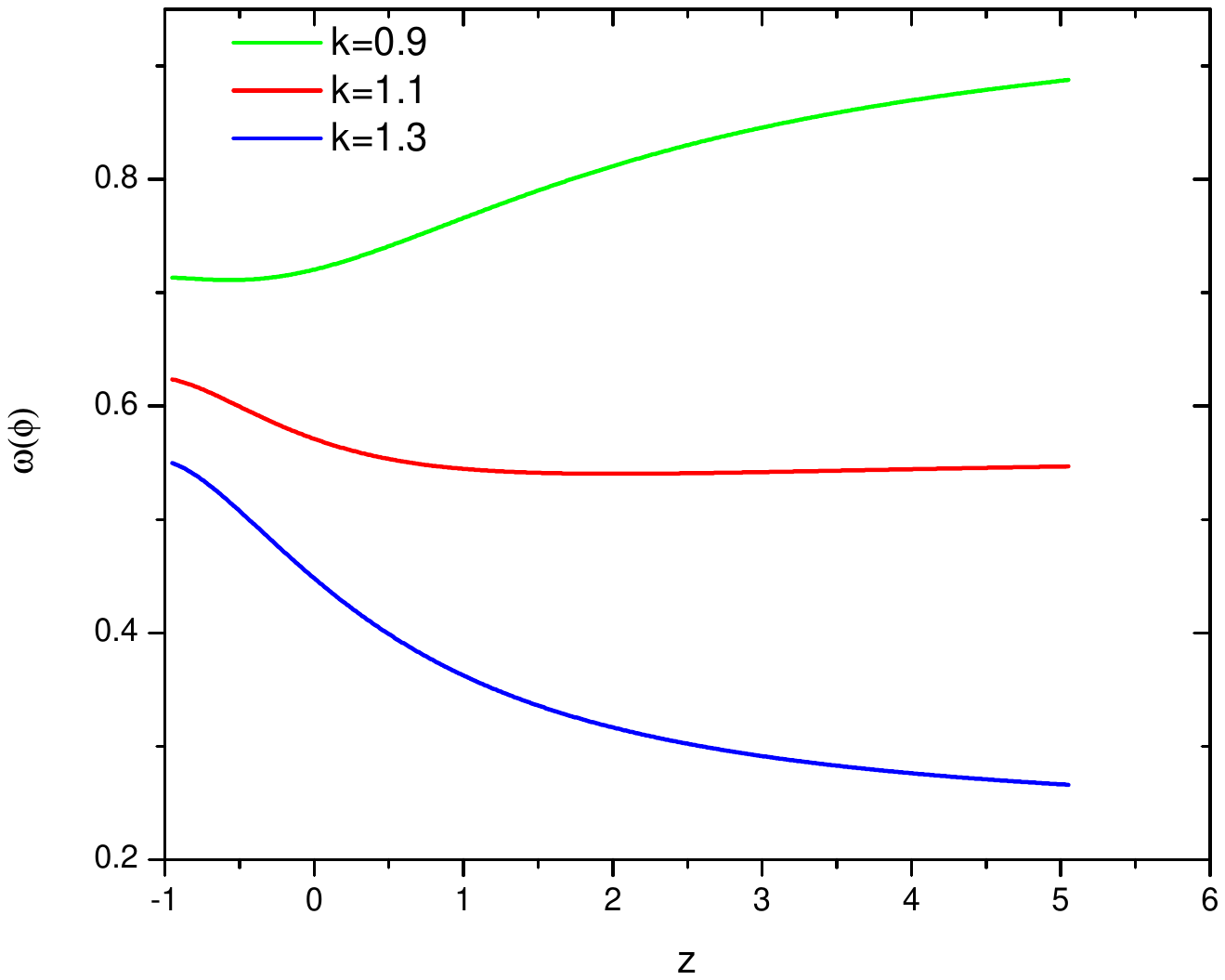}
\caption{ Evaluation of BD parameter $\omega(\varphi)$ as function of red shift with z . Here, the unified dark fluid parameter $\zeta$ is taken to be $0.00172$.}
\label{Fig5}
\end{figure}
%*********************************
%*********************************
\subsection{Brans-Dicke parameter}
In the GBD theory, the parameter $\omega$ is considered as a function of the BD scalar field. We have also derived the general expression for $\omega$ which depends on an assumed relationship between the rest energy density and proper pressure, the Hubble rate and the scalar field. It is to note that, in order to simulate the presence of a dark sector of the Universe, we have considered a unified dark fluid EoS which takes into account of both the dark energy and dark matter. With the assumed scale factor in the model, the expression of the BD parameter can be written as, 

\begin{eqnarray}\label{eq:32}
\omega(H) = \left(\frac{\dot{H}}{H}+3{H}\right)^{-2}\Box[-\dfrac{\rho+p}{\varphi}-\frac{\ddot{H}}{H}+6\dot{H}\nonumber\\+\dfrac{54}{(k+2)^{2}}H^{2}\Box]
\end{eqnarray}

Tripathy et al. \cite{Tripathy15}, have shown the behavior of the BD parameter with the power law and exponential expansion and observed that the anisotropic parameter only affects the non-evolving part of the BD parameter. In some other works, Tripathy and collaborators \cite{Pradhan2022, Tripathy2023} have shown that, the anisotropic parameter affects the non-evolving part of the BD parameter when the deceleration parameter comes out to be constant quantity. Of course, this deceleration parameter dependence of the BD parameter is quite model dependent \cite{Tripathy2023}. In the present case, one may observe that, it is not possible to separate out the evolving and non-evolving parts. In fact, the evolution of the BD parameter is more involved in the present model. Here we have presented graphically the evolutionary behaviour of the BD parameter for some representative values of the anisotropic parameter in FIG. 5. In order to obtain the plot, we fix the values of the UDF parameter as $\zeta=0.00172$. The effect of the anisotropic parameter on the evolutionary aspect of the BD parameter is clearly visible. Moreover, the BD parameter is more sensitive to the choice of the anisotropic parameter. It is seen that, for  a choice of $k=0.9$, the BD parameter $\omega(\varphi)$  decreases gradually from an early epoch to the late epoch. However, for higher values of the anisotropic parameter such as  $k=1.1$ and $k=1.3$, $\omega(\varphi)$ increases slowly from an early epoch to late epoch. At a given epoch, $\omega(\varphi)$ has a lower value for higher cosmic anisotropy. It is also interesting that, the values of $\omega(\varphi)$ at an initial epoch and at an final epoch vary widely with the choices of the anisotropic parameter. The predicted values of the BD parameter at the present epoch for different choices of the anisotropic parameter are given in Table-1.

\subsection{Self interacting potential}
The self interacting potential $V(\varphi)$, for the space-time \eqref{eq:4} and scale factor \eqref{eq:21} can be calculated as, \\
\\
\begin{strip}
\begin{eqnarray} \label{eq:33}
V(\varphi) = 2\varphi \left[\frac{9(2k+1)H^{2}}{(k+2)^{2}}-\frac{\rho}{\varphi}-\frac{\omega(\varphi)}{2}\left(-\frac{\dot{H}}{H}-3{H}\right)^{-2}+3H\left(\frac{\dot{H}}{H}-3{H}\right)\right]
\end{eqnarray}
\end{strip}

Graphically the behaviour of $V(\varphi)$ for the representative values of cosmic anisotropy is represented in FIG. 6 using the UDF parameter $\zeta=0.00172$. In general, the self interacting potential is negative for all the choices of the anisotropic parameter. The evolutionary behaviour shows that the self interacting potential increases from some large negative value to  almost vanish at the late time of  the evolution. The self interacting potential is also sensitive to the choice of $k$. At an initial epoch, for higher $k$, the curve of  $V(\varphi)$ remains atop of those with lower values of $k$. However, at the present epoch and at late epochs, all the curves of  $V(\varphi)$ overlap and vanish at late times. In Table-1, we have presented the predicted values of the self interacting potential required for the viability of the model for different values of the anisotropic parameter. 
\begin{figure}[!htp]
\centering
\includegraphics[scale=0.35]{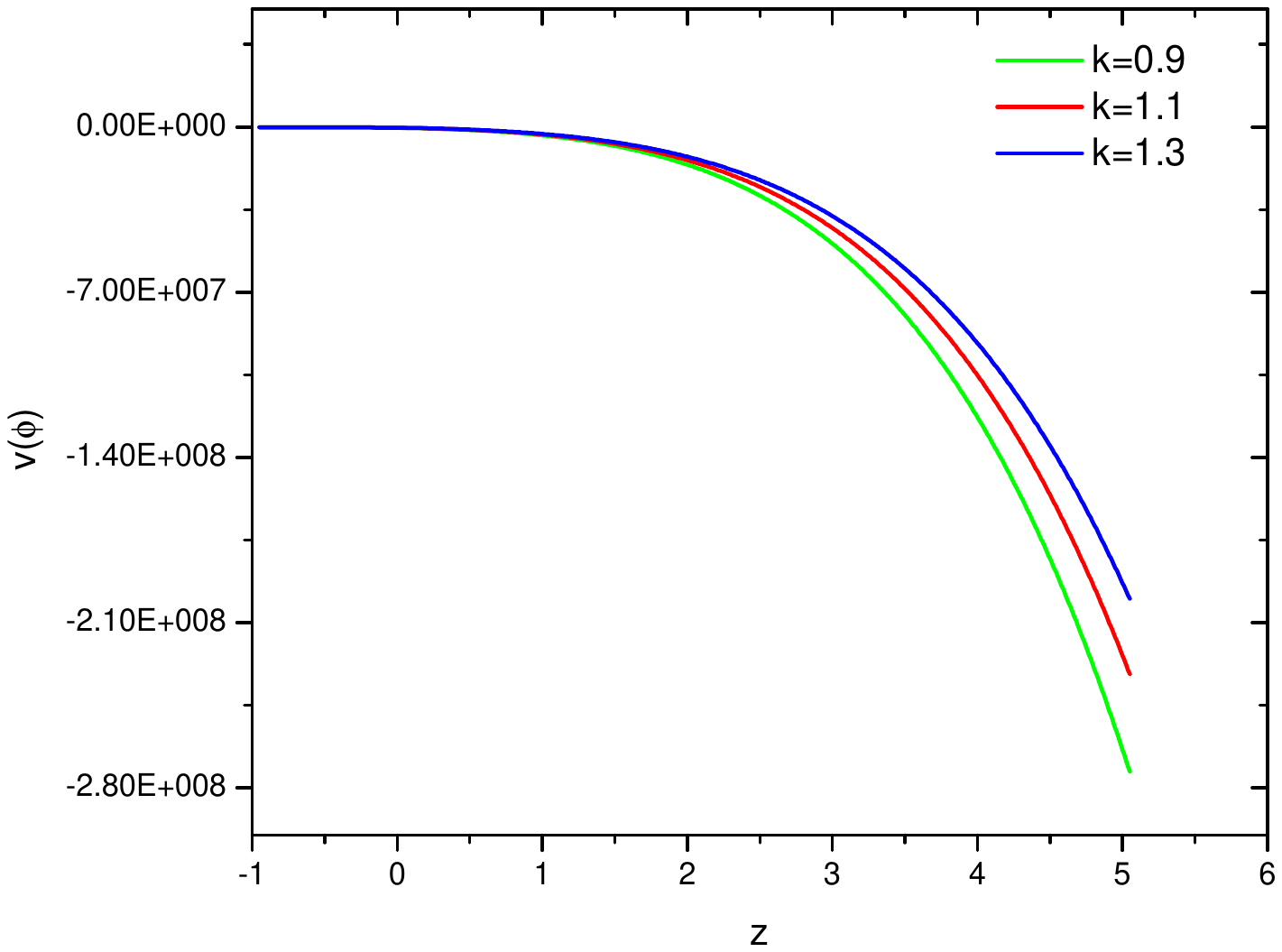}
\caption{Evolution of the self interacting potential $V(\varphi)$ is shown as a function of red shift $z$}
\label{Fig6}
\end{figure}
\begin{figure}[!htp]
\centering
\includegraphics[scale=0.35]{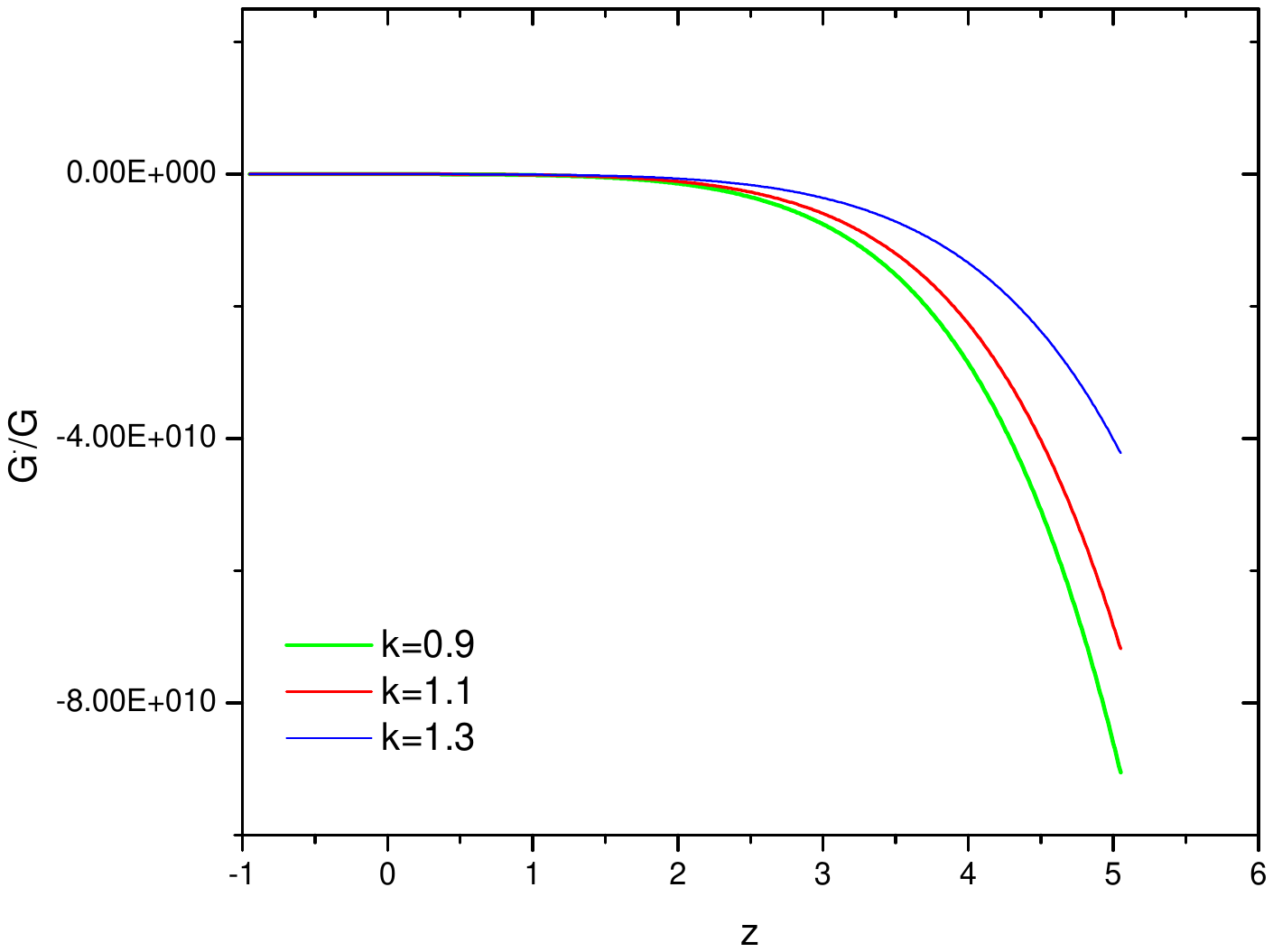}
\caption{ $\frac{\dot{G}}{G}$ as function of redshift $z$. The unified dark fluid parameter $\zeta$ is taken to be $0.00172$.}
\label{Fig6}
\end{figure}
\subsection{Variation of Newtonian Gravitational Constant $G$}
In generalized BD theory, the Newtonian gravitational constant $G(\varphi)$ is expressed as \cite{Nordtvedt70, Benkenstein1978, Alimi1996, Tripathy2020}
\begin{equation} \label{eq:34}
G(\varphi)=\frac{4+2\omega(\varphi)}{\varphi(3+2\omega(\varphi))}.
\end{equation}
The time variation of the Newtonian gravitational constant is obtained as
\begin{equation} \label{eq:35}
\frac{\dot{G}}{G}=-\frac{2\dot {\omega}(\varphi)}{(4+2\omega(\varphi))(3+2\omega(\varphi))}-\frac{\dot{\varphi}}{\varphi},
\end{equation}
which may be reduced to $\frac{\dot{G}}{G}=-\frac{\dot{\varphi}}{\varphi}$ for a non-evolving BD parameter $\omega$. In other words, the time evolution of the Newtonian Gravitational constant comes from the time evolution of the BD scalar field. However, in the present case with a dynamical BD parameter, we have some extra contribution coming from the dynamical nature of the BD parameter. It is worth to mention here that the Newtonian gravitational constant $G$ witnesses a very small variation whose measurement from different observations usually involve large errors. Different estimates for the variation of the $G$ include that of $\frac{\dot{G}}{G}(z=0)\simeq (2\pm 4) \times 10^{-12}~yr^{-1}$ from the Viking Lander Ranging \cite{Viking}, $\frac{\dot{G}}{G}(z=0) < 4 \times 10^{-10}~yr^{-1}$ from the gravochemical heating process \cite{Gravo} and $\frac{\dot{G}}{G}(z=0) \simeq (1.10\pm 1.07) \times 10^{-11}~yr^{-1}$ from the observations of double Neutron star binary \cite{Damour1991}. In FIG.7, we have plotted $\frac{\dot{G}}{G}$ as a function of redshift $z$ for $\zeta=0.00172$. It may be observed that, the variation of $G$ is prominent at an early epoch and becomes negligible at late times. The variation of $G$ is also affected by the choice of the anisotropic parameter. At least at late times, the variation in $G$ is substantial as compared to that for lower values of the anisotropic parameter.

\begin{table}
\caption{ The predicted values of Hubble rate ($H_0$), Deceleration parameter ($q_0$), Brans-Dicke parameter ($\omega_0$), Self interacting potential ($\omega_0$), Transition red shift ($z_0$), at the present epoch.}
\centering
\begin{tabular}{c|c}
\hline
\hline
Parameters & Predicted value \\
\hline
$H_0$&64.57\\
$q_0$&-0.32\\
$\omega_0(k=0.9)$&0.72\\
$\omega_0(k=1.1)$&0.571\\
$\omega_0(k=1.3)$&0.448\\
$V_0(k=0.9)$&-343982.6\\
$V_0(k=1.1)$&-320553.2\\
$V_0(k=1.9)$&-302539.8\\
$z_{t}$&0.68\\
\hline
\end{tabular}
\end{table}
%\newpage
\section{Final Remarks}

In this paper, we have presented the dark energy cosmological model in generalized  BD theory. We have assumed an anisotropic metric in the form of Bianchi type I space-time. Since our intention is to study some more details on the late time cosmic speed up phenomena, we have used the unknown form of energy, the concept of the dark energy as the unified dark fluid EoS. In order to frame the cosmological model, we have considered a scale factor that allows the deceleration parameter to vary with time showing a signature flipping behaviour. Basing upon a tabulated observational $H(z)$ data we constrained the parameters of the scale factor and that of the deceleration parameter. Some of the salient features of the model are:

\begin{itemize}
\item The observational constraints obtained on the model parameters are, $\delta=36.67\pm 9.58 $, $ \beta=27.90\pm 7.09$, $\gamma=1.58\pm 0.17$. With these constraints, the Hubble parameter shows an increasing trend with an increase in redshift signifying an accelerating Universe at the present epoch. The present value of the Hubble parameter as obtained from the present model is $H_0=64.57 \pm 16.67 km~s^{-1}Mpc^{-1}$ which is in close conformity with other recent estimates. 

\item The deceleration parameter for the present model shows a cosmic transit behaviour with an early deceleration to late time acceleration. The cosmic transit occurs at a redshift $z_t=0.68$. At the present epoch, the deceleration parameter is found to be $q_0=-0.32 \pm 0.42$. These estimates shows an accelerating expansion of the Universe at the present epoch. 
    
\item A linear EoS is considered in the present work in the form of the unified dark fluid that has a combined effect of dark energy and dark matter sector of the Universe commonly known as dark degeneracy. The parameters of the UDF EoS are considered on the basis of providing suitable mechanical stability in the cosmic fluid given through the speed of sound within the fluid. The effective equation of state parameter is observed to depend on the choice of the UDF parameters and shows a gradual evolution from an early positive value to late time negative values. From a simple analysis, we inferred that, the present model may favour a phantom like phase.

\item Using the constrained model along with the concept of unified dark fluid, different aspects of the generalized BD theory have been determined. The evolution of the BD scalar field is shown to have a decreasing trend with cosmic time. At the present time, it decreases to a vanishingly small value.

\item The BD parameter is assumed to vary with the scalar field and have a dynamical trajectory. The evolution of the BD parameter is found to be more sensitive to the choice of the anisotropic parameter. Usually, for constant deceleration parameter, it is possible to separate out the evolving part of the BD parameter from its non-evolving part and the anisotropic parameter can affect only its non-evolving part. However in the present cosmic transit model, the evolutionary behaviour of  BD parameter is more involved and is controlled by the anisotropic parameter as can be seen from FIG.5. The interesting aspect is, while for low values of the anisotropic parameter i.e. $k=0.9$, it decreases gradually with cosmic time, for higher values of the anisotropic parameter, $k>1$, it shows an increasing  trend.  Interestingly, the trajectory of the BD parameter for $k=1.3$ appears to be a mirror image of that  for $k=0.9$. At a given epoch, the BD parameter assumes a lower value for larger cosmic anisotropy. 

\item The evolutionary behaviour of the self interacting potential and that for variation of the Newtonian gravitational constant are alike and are mostly anisotropy dependent. The self interacting potential increase from a higher negative value; however from different time zone for the representative value of the anisotropic parameter. It is worthy to note here that at late times the self interacting potential vanishes irrespective of the value of the anisotropic parameter. $\frac{\dot{G}}{G}$ also shows the similar behaviour and vanishes at late time of the evolution. At least at late times, the variation in $G$ is substantial as compared to that for lower values of the anisotropic parameter.
\end{itemize}

\section*{Acknowledgement}
SKT and BM thank IUCAA, Pune (India) for providing support during an academic visit where a part of this work is carried out.

\section*{Conflict of interest:}
The authors have stated explicitly that there are no conflicts of interest in connection with this article.

% ===============================

\end{document}